\documentclass[11pt]{article}
\usepackage{amssymb}

\usepackage[totalwidth=460truept,totalheight=600truept]{geometry}
\usepackage{latexsym,amssymb,amsmath,graphicx}
\usepackage[T1]{fontenc}

\def\theequation{\arabic{section}.\arabic{equation}}

\renewcommand{\theequation}{\thesection.\arabic{equation}}
\global\arraycolsep=1truept

\begin{document}

\hfill IFUP-TH 2008/16

\vskip 1.4truecm

\begin{center}
{\huge \textbf{Weighted Power Counting, Neutrino Masses}}

{\huge \textbf{\large \vskip .1truecm}}

{\huge \textbf{And\ Lorentz Violating}}

{\huge \textbf{\large \vskip .1truecm}}

{\huge \textbf{Extensions Of The Standard Model}}

\vskip 1.5truecm

\textsl{Damiano Anselmi}

\textit{Dipartimento di Fisica ``Enrico Fermi'', Universit\`{a} di Pisa, }

\textit{Largo Pontecorvo 3, I-56127 Pisa, Italy, }

\textit{and INFN, Sezione di Pisa, Pisa, Italy}

damiano.anselmi@df.unipi.it

\vskip 2truecm

\textbf{Abstract}
\end{center}

\bigskip

{\small We study the Standard-Model extensions that have the following
features: they violate Lorentz invariance explicitly at high energies; they
are unitary, local, polynomial and renormalizable by weighted power
counting; they contain the vertex }$(LH)^{2}${\small , which gives Majorana
masses to the neutrinos after symmetry breaking, and possibly four fermion
interactions; they do not contain right-handed neutrinos, nor other extra
fields. We study the simplest CPT invariant Standard-Model extension of this
type in detail and prove the cancellation of gauge anomalies. We investigate
the low-energy recovery of Lorentz invariance and comment on other types of
extensions.}

\vskip 1truecm

\vfill\eject

\section{Introduction}

\setcounter{equation}{0}

Lorentz symmetry is a basic ingredient of the Standard Model of particle
physics. However, several authors have argued that at high energies Lorentz
symmetry and CPT could be broken \cite{colective,kostelecky,colective2}. The
problem has been attracting a lot of interest. In the power-counting
renormalizable sector the parameters of the Lorentz violating Standard-Model
extension \cite{kostelecky} have been tested with great precision \cite
{koste} and found in agreement with Lorentz invariance. However, if Lorentz
symmetry were explicitly violated at very high energies our understanding of
Nature would change substantially.

If we do not assume Lorentz invariance, yet demand unitarity, we discover
that the set of local renormalizable theories is considerably larger than
usual. Interactions that are not renormalizable by ordinary power counting
become renormalizable in a more general framework, called ``weighted power
counting'' \cite{renolor}, where space and time have different weights.
Because of unitarity no terms containing higher time derivatives are
allowed. Yet, since Lorentz symmetry is violated, terms containing higher
space derivatives can be present. Quadratic terms of this type can improve
the behaviors of propagators at large momenta and allow us to renormalize
interactions that otherwise would be non-renormalizable, saving
polynomiality. At the same time, weighted power counting ensures that
renormalization does not regenerate terms containing higher time
derivatives. Clearly, the high-energy behavior of the theory is modified in
an essential way. For this reason, the Lorentz violation we are talking
about cannot be spontaneous, but must be explicit.

It is interesting to inquire what physics beyond the Standard Model emerges
in this approach, assuming that Lorentz symmetry is violated by terms of
higher dimensions and restored at low energies. In non-gauge theories
several types of Lorentz breakings are allowed \cite{renolor,confnolor}. The
presence of gauge interactions puts more severe restrictions \cite
{LVgauge1suA,LVgauge1suAbar}, for example it privileges the Lorentz breaking
where spacetime is split into space and time. Lorentz invariance can be
violated preserving or not preserving CPT \cite{cpt}. We concentrate the
major part of our attention on the ``minimal breaking'' of Lorentz symmetry,
which preserves CPT\ and invariance under space rotations. We construct CPT\
invariant Standard-Model extensions that contain two scalar-two fermion
interactions and four fermion interactions. In particular, the models
contain the vertex $(LH)^{2}$ \cite{strumia}, which gives Majorana masses to
the neutrinos after symmetry breaking. No right-handed neutrinos, nor other
extra fields, are present.

Once Lorentz symmetry is violated at high energies, its low-energy recovery
is not automatic, because renormalization makes the low-energy parameters
run independently. We have to advocate a fine tuning that relates such
parameters in a suitable way. It is not apparent how to justify this
fine-tuning, unless the Lorentz invariant surface is RG\ stable \cite
{nielsen}.

The relation between neutrino oscillations and Lorentz violation has been
widely explored in the literature. Cohen and Glashow formulated a theory of
``very special relativity'' \cite{cohen2} according to which the exact
symmetry group of Nature includes space-time translations and a proper
subgroup of the Lorentz group. In this framework, they proposed a
Lorentz-violating origin for lepton-number conserving neutrino masses \cite
{cohen1}, without need for right-handed neutrinos. In the Cohen-Glashow
approach the neutrinoless double beta decay is forbidden. Kosteleck\'{y} and
Mewes \cite{mewes} proposed that the observed data about neutrino
oscillations be explained by Lorentz and CPT violations rather than mass
differences. They studied the neutrino behavior in the framework of the
minimal Standard-Model extension of Colladay and Kosteleck\'{y} \cite
{kostelecky} without neutrino masses. Their investigation was extended by
Katori, Kosteleck\'{y} and Tayloe \cite{katori}, who showed that when the
neutrino mass terms are included more data about neutrino oscillations can
be accommodated. Another mechanism to explain neutrino oscillations without
neutrino masses, due to Klinkhamer, is based on a Fermi-point-splitting
method, suggested by an analogy with condensed-matter physics \cite{klink}.

In this paper our main concern is to show that the violation of Lorentz
symmetry allows us to extend the Standard Model to include two scalar-two
fermion vertices, as well as four fermion vertices, in a renormalizable way,
and that to achieve this goal it is not necessary to violate CPT. Closure
under renormalization requires that several other vertices and quadratic
terms be present. The model predicts a departure from the Standard Model
starting from energies of the order of $\sim $10$^{14}$GeV and a completely
new kind of UV behavior. Further work is necessary to search for
experimentally detectable effects.

The paper is organized as follows. In section 2 we review the weighted power
counting. In section 3 we classify the Lorentz violating extensions of the
Standard Model, focusing on the CPT\ invariant solutions that contain two
scalar-two fermion vertices and four fermion vertices. In section 4 we write
the simplest model with such features in detail. In section 5 we prove that
the gauge anomalies of our extended model coincide with those of the
Standard Model, therefore they cancel out to all orders. Our analysis
provides also an alternative, general proof of the Adler-Bardeen theorem 
\cite{adler,adler2}. Section 6 contains our conclusions. In the appendix we
recall the form of the gauge-field propagator and the problem of spurious
subdivergences.

\section{Weighted power counting}

\setcounter{equation}{0}

The simplest framework to study Lorentz violations is to assume that the $d$%
-dimensional spacetime manifold $M=\mathbb{R}^{d}$ is split into the product 
$\hat{M}\times \bar{M}$ of two submanifolds, a $\hat{d}$-dimensional
submanifold $\hat{M}=\mathbb{R}^{\hat{d}}$, containing time and possibly
some space coordinates, and a $\bar{d}$-dimensional space submanifold $\bar{M%
}=\mathbb{R}^{\bar{d}}$, so that the $d$-dimensional Lorentz group $O(1,d-1)$
is broken to a residual Lorentz group $O(1,\hat{d}-1)\times O(\bar{d})$. The
formalism developed for the two-factor splitting of $M$ can be generalized
to treat the most general Lorentz violation. However, in ref.s \cite
{LVgauge1suA,LVgauge1suAbar} it is shown that the absence of certain
spurious divergences in Feynman diagrams selects a two-factor split with $%
\hat{d}=1$, which we are going to assume henceforth. This split is the
physically most interesting one, since it states that at high energies there
is no ``spacetime'', but just space and time. In this section we assume
separate C, P and T invariances. We later relax these assumptions to study
the Standard Model.

The partial derivative $\partial $ is decomposed as $(\hat{\partial},\bar{%
\partial})$, where $\hat{\partial}$ and $\bar{\partial}$ act on the
subspaces $\hat{M}$ and $\bar{M}$, respectively. Coordinates and momenta are
decomposed similarly. Consider a scalar theory with quadratic (Euclidean)
lagrangian 
\begin{equation}
\mathcal{L}_{\hbox{free}}=\frac{1}{2}(\hat{\partial}\varphi )^{2}+\frac{1}{%
2\Lambda _{L}^{2n-2}}(\bar{\partial}^{n}\varphi )^{2},  \label{free}
\end{equation}
where $\Lambda _{L}$ is an energy scale and $n$ is an integer $>1$. The
theory (\ref{free}) is invariant under the weighted rescaling 
\begin{equation}
\hat{x}\rightarrow \hat{x}\ \mathrm{e}^{-\Omega },\qquad \bar{x}\rightarrow 
\bar{x}\ \mathrm{e}^{-\Omega /n},\qquad \varphi \rightarrow \varphi \ 
\mathrm{e}^{\Omega (\text{\dj }/2-1)},  \label{scale}
\end{equation}
where \dj $=\hat{d}+\bar{d}/n$ is called ``weighted dimension''. Note that $%
\Lambda _{L}$ is not rescaled.

The interacting theory is defined as a perturbative expansion around (\ref
{free}). For the purposes of renormalization, the masses and other quadratic
terms can be treated perturbatively, since the counterterms depend
polynomially on them. Denote the ``weight'' of an object $\mathcal{O}$ by $[%
\mathcal{O}]$ and assign weights to coordinates, momenta and fields as
follows: 
\begin{equation}
\lbrack \hat{x}]=-1,\qquad [\bar{x}]=-\frac{1}{n},\qquad [\hat{\partial}%
]=1,\qquad [\bar{\partial}]=\frac{1}{n},\qquad [\varphi ]=\frac{\text{\dj }}{%
2}-1,\qquad [\psi ]=\frac{\text{\dj }-1}{2},  \label{weights}
\end{equation}
while $\Lambda _{L}$ is weightless.

The vertices of weight \dj\ are strictly renormalizable, those of weight
smaller than \dj\ super-renormalizable and those of weight greater than \dj\
non-renormalizable. A theory is renormalizable if it contains no
non-renormalizable vertices. This condition ensures also that the theory
does not contain higher time derivatives, which guarantees perturbative
unitarity.

Having decomposed the partial derivative $\partial $ as $(\hat{\partial},%
\bar{\partial})$, the gauge field has to be decomposed similarly. We write $%
A=(\hat{A},\bar{A})$, so the covariant derivative reads 
\begin{equation}
D=(\hat{D},\bar{D})=(\hat{\partial}+g\hat{A},\bar{\partial}+g\bar{A}).
\label{3a}
\end{equation}
where $g$ is the gauge coupling. Then, we have the weight assignments 
\begin{equation}
\lbrack g\hat{A}]=[\hat{\partial}]=1,\qquad [g\bar{A}]=[\bar{\partial}]=%
\frac{1}{n}.  \label{3c}
\end{equation}
The field strength is decomposed into 
\begin{equation}
\tilde{F}_{\mu \nu }\equiv F_{\hat{\mu}\bar{\nu}},\qquad \bar{F}_{\mu \nu
}=F_{\bar{\mu}\bar{\nu}}.  \label{3b}
\end{equation}
while $\hat{F}_{\mu \nu }=0$ at $\hat{d}=1$.

The BRST\ symmetry \cite{brs} is the same as usual, 
\[
sA_{\mu }^{a}=D_{\mu }^{ab}C^{b},\qquad sC^{a}=-\frac{g}{2}%
f^{abc}C^{b}C^{c},\qquad s\bar{C}^{a}=B^{a},\qquad sB^{a}=0,\qquad s\psi
^{i}=-gT_{ij}^{a}C^{a}\psi ^{j}, 
\]
etc. The simplest gauge-fixing is 
\begin{equation}
\mathcal{L}_{\text{gf}}=s\Psi ,\qquad \Psi =\bar{C}^{a}\left( -\frac{\lambda 
}{2}B^{a}+\mathcal{G}^{a}\right) ,\qquad \mathcal{G}^{a}\equiv \hat{\partial}%
\cdot \hat{A}^{a}+\zeta \left( \bar{\upsilon}\right) \bar{\partial}\cdot 
\bar{A}^{a}  \label{gf}
\end{equation}
where $\lambda $ is a dimensionless, weightless constant, $\bar{\upsilon}%
\equiv -\bar{\partial}^{2}/\Lambda _{L}^{2}$ and $\zeta $ is a polynomial of
degree $n-1$.

At $\hat{d}=1$ the field strength does not contain terms of the form $\hat{%
\partial}\hat{A}$, which however enter the gauge fixing. Once the Lagrange
multiplier $B^{a}$ is integrated out the gauge-fixed action contains $(\hat{%
\partial}\hat{A})^{2}$, which must have weight \dj . Therefore, the weight
of $\hat{A}$ is \dj $/2-1$. Then, (\ref{3c}) gives $[g]=2-$\dj $/2$. From $%
\tilde{F}$ we get $[\bar{\partial}]+[\hat{A}]=[\hat{\partial}]+[\bar{A}]$,
so we derive $[\bar{A}]$. In summary, 
\begin{equation}
\lbrack \hat{A}]=\frac{\text{\dj }}{2}-1,\qquad [\bar{A}]=\frac{\text{\dj }}{%
2}-2+\frac{1}{n},\qquad [\hat{F}]=\frac{\text{\dj }}{2},\qquad [\tilde{F}]=%
\frac{\text{\dj }}{2}-1+\frac{1}{n},\qquad [\bar{F}]=\frac{\text{\dj }}{2}-2+%
\frac{2}{n}.  \label{we}
\end{equation}

The gauge-field action 
\begin{equation}
\mathcal{S}_{0}=\int \mathrm{d}^{d}x\left( \mathcal{L}_{Q}+\mathcal{L}%
_{I}\right) \equiv \mathcal{S}_{Q}+\mathcal{S}_{I},  \label{s0}
\end{equation}
is the sum of two contributions: the quadratic terms $\mathcal{S}_{Q}$,
which are constructed with two field strengths and possibly covariant
derivatives; the vertex terms $\mathcal{S}_{I}$, which are constructed with
at least three field strengths, and possibly covariant derivatives.

Up to total derivatives the form of the quadratic part $\mathcal{L}_{Q}$ of
the lagrangian reads (in the Euclidean framework) \cite{LVgauge1suA} 
\begin{equation}
\mathcal{L}_{Q}=\frac{1}{4}\left\{ 2F_{\hat{\mu}\bar{\nu}}\eta (\bar{\Upsilon%
})F_{\hat{\mu}\bar{\nu}}+F_{\bar{\mu}\bar{\nu}}\tau (\bar{\Upsilon})F_{\bar{%
\mu}\bar{\nu}}+\frac{1}{\Lambda _{L}^{2}}(D_{\hat{\rho}}F_{\bar{\mu}\bar{\nu}%
})\xi (\bar{\Upsilon})(D_{\hat{\rho}}F_{\bar{\mu}\bar{\nu}})\right\} .
\label{l0}
\end{equation}
Here $\bar{\Upsilon}\equiv -\bar{D}^{2}/\Lambda _{L}^{2}$ and $\eta $, $\tau 
$ and $\xi $ are polynomials of degrees $n-1$, $2n-2$ and $n-2$,
respectively.

Finally, the gauge-fixed action reads 
\begin{equation}
\mathcal{S}=\int \mathrm{d}^{d}x\left( \mathcal{L}_{Q}+\mathcal{L}_{I}+%
\mathcal{L}_{\text{gf}}\right) \equiv \mathcal{S}_{0}+\mathcal{S}_{\text{gf}%
}.  \label{basis}
\end{equation}

In the appendix we recall the form of the gauge-field propagator and the
origin and disappearance of spurious subdivergences. The two physical
degrees of freedom have the dispersion relation 
\begin{equation}
E=\sqrt{\bar{k}^{2}\frac{\tau (\bar{k}^{2}/\Lambda _{L}^{2})}{\tilde{\eta}(%
\bar{k}^{2}/\Lambda _{L}^{2})}},  \label{disp}
\end{equation}
where $\tilde{\eta}=\eta +\xi \bar{k}^{2}/\Lambda _{L}^{2}$.

The renormalizability requirement can be refined to single out the
super-renormalizable theories. Observe that the gauge coupling is always
super-renormalizable in four dimensions, for $n>1$. Introduce a coupling $%
\bar{g}$ of non-negative weight $\kappa $ and demand that every vertex with,
say, $N$ legs be multiplied by $\lambda _{c}\bar{g}^{N-2}$, and that the
weight of $\lambda _{c}$ be non-negative. It is easy to see that the
counterterms generated by such vertices are multiplied by coefficients that
have the same structure, so no new counterterms are turned on by
renormalization and the theory is renormalizable.

The requirement can be further refined allowing different fields to have
different $\bar{g}$'s. For example, call $\bar{g}_{i}$, $i=1,2,3,$ the ones
of vectors, fermions and scalars, respectively. The most general lagrangian
has the weight structure \cite{LVgauge1suAbar} 
\begin{eqnarray}
\mathcal{L} &=&\frac{1}{\bar{\alpha}_{1}}\mathcal{L}_{1}(\bar{g}_{1}A)+\frac{%
1}{\bar{\alpha}_{2}}\mathcal{L}_{2}(\bar{g}_{2}\psi )+\frac{1}{\bar{\alpha}%
_{3}}\mathcal{L}_{3}(\bar{g}_{3}\varphi )+\frac{1}{\bar{a}_{3}}\mathcal{L}%
_{12}(\bar{g}_{1}A,\bar{g}_{2}\psi )  \nonumber \\
&&+\frac{1}{\bar{a}_{2}}\mathcal{L}_{13}(\bar{g}_{1}A,\bar{g}_{3}\varphi )+%
\frac{1}{\bar{a}_{1}}\mathcal{L}_{23}(\bar{g}_{2}\psi ,\bar{g}_{3}\varphi )+%
\frac{1}{\bar{\alpha}}\mathcal{L}_{123}(\bar{g}_{1}A,\bar{g}_{2}\psi ,\bar{g}%
_{3}\varphi ).  \label{mixed}
\end{eqnarray}
Here $\bar{\gamma}_{k}$, $k=1,2,3$, denote the couplings of minimum weight
between $\bar{g}_{i}$ and $\bar{g}_{j}$, where $k\neq i,j$. Instead, $\bar{g}
$ is the coupling of minimum weight among the $\bar{g}_{i}$'s. We have
defined $\bar{\alpha}_{i}=\bar{g}_{i}^{2}$, $\bar{a}_{i}=\bar{\gamma}%
_{i}^{2} $. In $A$ we have collectively included also ghosts and antighosts.
Every other parameter $\lambda $ contained in (\ref{mixed}) must have a
non-negative weight. The $\bar{g}_{i}$-factors appearing in formula (\ref
{mixed}) are mere tools to keep track of the weight structure. For example,
instead of $\bar{g}_{2}\psi $ we can have any $\bar{g}_{i}\psi $, as long as 
$[\bar{g}_{i}]\geq [\bar{g}_{2}]$. Similarly, the denominators $1/\bar{\alpha%
}_{i}$, $1/\bar{a}_{i}$ and $1/\bar{\alpha}$ are devices that lower the
weights of appropriate amounts.

The one-loop counterterms generated by (\ref{mixed}) have the weight
structure 
\begin{equation}
\Delta _{1}\mathcal{L}(\bar{g}_{1}A,\bar{g}_{2}\psi ,\bar{g}_{3}\varphi ),
\label{mixec}
\end{equation}
while at $L$ loops there is an additional factor of $\bar{\alpha}^{L-1}$. A
simplified version of the theory can be obtained dropping vertices and
quadratic terms of (\ref{mixed}) that are not contained in (\ref{mixec}),
because renormalization is unable to generate them back. Of course, the
quadratic terms that are crucial for the behaviors of propagators, and the
vertices related to them by covariantization, must be kept in any case.

Every $\mathcal{L}$ on the right-hand side of (\ref{mixed}) must be
polynomial in the fields and parameters, which happens if 
\begin{equation}
4-\frac{4}{n}-2\kappa _{1}<\text{\dj },\qquad 1-2\kappa _{2}<\text{\dj }%
,\qquad 2-2\kappa _{3}<\text{\dj .}  \label{h}
\end{equation}
having written $[\bar{g}_{i}]=\kappa _{i}\geq 0$. Compatibility with the
covariant structure demands 
\begin{equation}
\lbrack g]\geq [\bar{g}_{1}],\qquad [g\bar{g}_{1}]\geq [\bar{g}%
_{2}^{2}],\qquad [g\bar{g}_{1}]\geq [\bar{g}_{3}^{2}],  \label{tu}
\end{equation}
namely 
\begin{equation}
\kappa _{1}\leq 2-\frac{\text{\dj }}{2},\qquad \kappa _{2,3}\leq 1+\frac{%
\kappa _{1}}{2}-\frac{\text{\dj }}{4}.  \label{2h}
\end{equation}
Moreover, we must have $d\geq 4$ to ensure the absence of IR\ divergences in
Feynman diagrams, and $\hat{d}=1$, $d=$even, $n=$odd, together with some
extra restrictions \cite{LVgauge1suAbar}, to ensure the absence of spurious
subdivergences. The set of extra restrictions we are interested in is 
\begin{equation}
\text{\dj }\leq 2,\qquad \kappa _{1}>2-\frac{1}{n}-\frac{\text{\dj }}{2}%
,\qquad \kappa _{2}\geq 1-\frac{\text{\dj }}{2}.  \label{spurio}
\end{equation}
The absence of spurious subdivergences ensures the locality of counterterms.

Finally, the time-derivative structure of the theory is under control \cite
{LVgauge1suAbar}. No terms with more than two time derivatives are allowed
by weighted power counting.

\section{Lorentz violating extensions of the Standard Model}

\setcounter{equation}{0}

In this section we search for Lorentz violating renormalizable extensions of
the Standard Model. In particular, we investigate the existence of ``more
economic'' alternatives to it. Three interesting problems come to mind: $i$)
give masses to the gauge bosons without Higgs fields; $ii$) give masses to
the left-handed neutrinos without introducing right-handed neutrinos or
other extra fields, and without violating CPT; $iii$) include proton decay.
Option $i$)\ is not viable, because the Proca versions of our theories do
not have well-behaved propagators at infinity \cite{LVgauge1suA}. Instead,
options $ii$) and $iii$) are allowed, together with other types of Lorentz
violating extensions.

Since the Standard Model violates parity and time reversal (assuming that
CPT\ is exact) we must start from the proper, orthochronous Lorentz group $%
SO_{+}(1,3)$. The minimal Lorentz breaking preserves full invariance under
rotations and CPT. In four dimensions we have 
\begin{equation}
\text{\dj }=1+\frac{3}{n}.  \label{pic}
\end{equation}
Inserting (\ref{pic}) in (\ref{h}), (\ref{2h}) and (\ref{spurio}) we get 
\begin{eqnarray}
n=\text{odd}\geq 3,\qquad &&\frac{3}{2}-\frac{5}{2n}<\kappa _{1}\leq \frac{3%
}{2}-\frac{3}{2n},\qquad \frac{1}{2}-\frac{3}{2n}\leq \kappa _{2},  \nonumber
\\
\frac{1}{2}-\frac{3}{2n} &<&\kappa _{3},\qquad \kappa _{2,3}\leq \frac{3}{4}-%
\frac{3}{4n}+\frac{\kappa _{1}}{2}.  \label{a}
\end{eqnarray}
Solutions exist for every odd $n$. The simplest models are those that have
the smallest values of $n$ and the largest values of $\kappa _{1,2,3}$.

A two scalar-two fermion vertex, which has the form $(\bar{g}_{2}^{2}\bar{g}%
_{3}^{2}/\bar{a}_{1})\varphi ^{2}\overline{\psi }\psi $, is renormalizable
if its weight is not greater than \dj , namely if 
\begin{equation}
\kappa _{2,3}\leq 1-\frac{3}{2n}.  \label{2s2f}
\end{equation}
A four fermion vertex is renormalizable if 
\begin{equation}
\kappa _{2}\leq \frac{1}{2}-\frac{3}{2n}.  \label{4f}
\end{equation}
Combining (\ref{a}), (\ref{2s2f}) and (\ref{4f}) we get 
\begin{equation}
n=\text{odd}\geq 3,\qquad \frac{3}{2}-\frac{5}{2n}<\kappa _{1}\leq \frac{3}{2%
}-\frac{3}{2n},\qquad \kappa _{2}=\frac{1}{2}-\frac{3}{2n},\qquad \frac{1}{2}%
-\frac{3}{2n}<\kappa _{3}\leq 1-\frac{3}{2n}.  \label{tutto}
\end{equation}
Again, solutions exist for every odd $n\geq 3$.

\bigskip

Consider options $ii$) and $iii$). Candidate neutrino mass terms such as 
\begin{equation}
a^{\mu }\bar{\psi}_{L}\gamma _{\mu }\psi _{L},\qquad \psi _{L}^{\alpha
}M_{\alpha \beta }\psi _{L}^{\beta }+\text{h.c.},  \label{cpt}
\end{equation}
where $a^{\mu }$ is a constant vector, $\alpha ,\beta $ are Lorentz indices
and $M_{\alpha \beta }$ is a constant matrix, violate either CPT, or
hermiticity, or hypercharge conservation. The hypercharge of $\psi
_{L}^{\alpha }\psi _{L}^{\beta }$ can be compensated by two Higgs fields. It
is well-known that the unique vertex with two Higgses and two leptons is 
\cite{strumia} 
\begin{equation}
(LH)^{2}\equiv \sum_{a,b=1}^{3}Y_{ab}\ \varepsilon _{ij}L_{i}^{\alpha
a}H_{j}\ \varepsilon _{\alpha \beta }\ \varepsilon _{kl}L_{k}^{\beta b}H_{l}+%
\text{h.c.,}  \label{lh2}
\end{equation}
where $H_{i}$ denotes the complex Higgs doublet and $L_{i}^{a}=(\nu
_{L}^{a},\ell _{L}^{a})$ is the lepton doublet. The indices $a$ and $b$
label the generations and $Y_{ab}$ are constants.

The vertex (\ref{lh2}) gives Majorana masses to the neutrinos after symmetry
breaking, but is not renormalizable by ordinary power counting. Normally, it
is introduced with the minimal seesaw mechanism \cite{seesaw} as an
effective vertex obtained integrating out right-handed neutrinos $\nu _{R}$
with large Majorana masses, Yukawa-coupled to $L$ and $H$. Alternative
seesaw mechanisms \cite{altseesaw} postulate the existence of intermediate
fermionic or scalar $SU(2)_{L}$ triplets. Instead, here we are interested in
Lorentz violating models that contain the vertex (\ref{lh2}) at the
fundamental level.

We know that $n$ must be odd. Another reason why even $n$'s are excluded is
CPT\ invariance. Indeed, if $n$ is even, $n=2k$, no fermionic quadratic term 
$\sim \bar{L}\bar{\partial}^{2k}L$, which is crucial for renormalizability,
is allowed. Even if Lorentz invariance is maximally violated, $SU(2)\times
U(1)$ invariant quadratic terms such as 
\begin{equation}
b^{\mu }\bar{L}\gamma _{\mu }\bar{\partial}^{2k}L  \label{n=2}
\end{equation}
are forbidden by CPT invariance or hermiticity.

The simplest choice is $n=3$. Then \dj $=2$ and (\ref{a}) and (\ref{2s2f})
become 
\begin{equation}
\frac{2}{3}<\kappa _{1}\leq 1,\qquad 0\leq \kappa _{2}\leq \frac{1}{2}%
,\qquad 0<\kappa _{3}\leq \frac{1}{2}.  \label{p1}
\end{equation}
On the other hand, (\ref{4f}) tells us that a four fermion vertex is
renormalizable only if 
\begin{equation}
\kappa _{2}=0.  \label{p2}
\end{equation}
The simplest choices for option $ii$) are 
\begin{equation}
\text{\dj }=2\qquad n=3,\qquad \kappa _{1}=1,\qquad \kappa _{2}=\kappa _{3}=%
\frac{1}{2}.  \label{sm2}
\end{equation}
The simplest choices for option $iii$) are 
\begin{equation}
\text{\dj }=2,\qquad n=3,\qquad \kappa _{1}=1,\qquad \kappa _{2}=0,\qquad
\kappa _{3}=\frac{1}{2}.  \label{sm3}
\end{equation}
The solution (\ref{sm2}) can be obtained from (\ref{sm3}) switching some
vertices off. Observe that $\bar{g}_{1}=g$ in both cases.

In the remainder of this section we describe the structure of such
solutions. First we focus on the terms that do not contain the tensor $%
\varepsilon _{\bar{\mu}\bar{\nu}\bar{\rho}}$.

The quadratic part of the lagrangian is the sum of $\mathcal{S}_{0}$ (\ref
{s0}) for the gauge fields, plus 
\[
\mathcal{L}_{\text{kin }H\psi }=\bar{\psi}_{L}(\hat{D}\!\!\!\!\slash+P_{1}(%
\bar{\Upsilon})\bar{D}\!\!\!\!\slash)\psi _{L}+\bar{\psi}_{R}(\hat{D}\!\!\!\!%
\slash+P_{1}^{\prime }(\bar{\Upsilon})\bar{D}\!\!\!\!\slash)\psi _{R}+|\hat{D%
}H|^{2}+H^{\dagger }P_{3}(\bar{\Upsilon})H, 
\]
for the fermions and Higgs field $H$, where the $P_{i}$'s are polynomials of
degree $i$. The vertices have the form 
\begin{equation}
\frac{\bar{g}_{1}^{q+r}\bar{g}_{2}^{2s}\bar{g}_{3}^{t}}{\bar{g}^{\prime 2}}%
\hat{D}^{k}\bar{D}^{m}\tilde{F}^{q}\bar{F}^{r}\psi ^{2s}H^{t},  \label{vv}
\end{equation}
where $\bar{g}^{\prime }$ is the $\bar{g}$ of minimum weight among those
appearing in the vertex, and $H^{t}$ includes both powers of $H$ and $%
H^{\dagger }$. The weight of (\ref{vv}) must not be larger than \dj . We
find 
\begin{equation}
k+\frac{m+4q+2r}{3}+s(1+2\kappa _{2})+\frac{t}{2}\leq 2+2\kappa ^{\prime },
\label{d}
\end{equation}
where $\kappa ^{\prime }=1$ if $s=t=0$, $\kappa ^{\prime }\leq 1/2$
otherwise.

Let us start from the pure gauge sector. We find $3k+m+4q+2r\leq 12$. By CPT
and rotational invariance, $k$ and $q$ must have the same parity. We have
only $k=2$, $q=0$, or $k=q=1$, or $k=0$, $q\leq 2$, which gives the vertices 
\begin{eqnarray*}
&&g\hat{D}^{2}\bar{F}^{3},\qquad g\hat{D}\bar{D}\tilde{F}\bar{F}^{2},\qquad g%
\bar{D}^{2}\tilde{F}^{2}\bar{F},\qquad g^{2}\tilde{F}^{2}\bar{F}^{2}, \\
&&g\bar{D}^{6}\bar{F}^{3},\qquad g^{2}\bar{D}^{4}\bar{F}^{4},\qquad g^{3}%
\bar{D}^{2}\bar{F}^{5},\qquad g^{4}\bar{F}^{6}.
\end{eqnarray*}
Here and in the formulas below we list only the vertices with the largest
numbers of legs and derivatives.

Next, consider the sector containing scalar fields, but no fermions. The
maximum number of scalar legs is 6. We have the vertices 
\begin{eqnarray*}
&&\bar{g}^{4}H^{6},\qquad \bar{g}^{2}\bar{D}^{2}H^{4},\qquad \bar{g}g^{2}%
\bar{F}^{2}H^{3},\qquad \bar{g}g\bar{D}^{2}\bar{F}H^{3},\qquad \bar{g}\bar{D}%
^{4}H^{3},\qquad g^{3}\bar{F}^{3}H^{2}, \\
&&g^{2}\bar{D}^{2}\bar{F}^{2}H^{2},\qquad g\bar{D}^{4}\bar{F}H^{2},
\end{eqnarray*}
where $\bar{g}\equiv \bar{g}_{3}$.

Four fermion terms exist only for $\kappa _{2}=0$, as already mentioned, and
must have $k=m=q=r=0$, so their form is simply 
\[
\bar{g}_{2}^{2}\bar{\psi}\psi \bar{\psi}\psi . 
\]
These are the only strictly-renormalizable vertices contained in our model.
Switching them off turns solution (\ref{sm3}) into solution (\ref{sm2}).

The vertices containing two fermions are 
\[
\bar{g}^{2}H^{2}\bar{\psi}\psi ,\qquad \bar{g}\bar{D}H\bar{\psi}\psi ,\qquad
g\bar{D}\bar{F}\bar{\psi}\psi . 
\]

Now we consider the CPT\ invariant terms that do contain the tensor $\bar{%
\varepsilon}\equiv \varepsilon _{\bar{\mu}\bar{\nu}\bar{\rho}}$, which have
not been studied in ref.s \cite{LVgauge1suA,LVgauge1suAbar}. Since $\bar{%
\varepsilon}$ can be converted into $\gamma $ matrices, the terms containing
fermions have already been covered, so we can study objects of the form 
\[
\frac{\bar{g}_{1}^{q+r}\bar{g}_{3}^{t}}{\bar{g}^{\prime 2}}\bar{\varepsilon}%
\hat{D}^{k}\bar{D}^{m}\tilde{F}^{q}\bar{F}^{r}H^{t}, 
\]
where $k+q$ and $m+q$ must be odd, by CPT\ and rotational invariance.
Proceeding as before, we find 
\begin{eqnarray}
&&\bar{\varepsilon}\hat{D}^{2}\tilde{F}\bar{F},\qquad \bar{\varepsilon}\hat{D%
}\bar{D}\tilde{F}^{2},\qquad g\bar{\varepsilon}\tilde{F}^{3},\qquad \bar{%
\varepsilon}\hat{D}\bar{D}^{5}\bar{F}^{2},\qquad \bar{\varepsilon}\bar{D}^{6}%
\tilde{F}\bar{F},\qquad  \nonumber \\
&&g\bar{\varepsilon}\hat{D}\bar{D}^{3}\bar{F}^{3},\qquad g\bar{\varepsilon}%
\bar{D}^{4}\tilde{F}\bar{F}^{2},\qquad g\bar{\varepsilon}\hat{D}\bar{D}\bar{F%
}H^{2},\qquad g\bar{\varepsilon}\bar{D}^{2}\tilde{F}H^{2},  \nonumber \\
&&g^{2}\bar{\varepsilon}\hat{D}\bar{D}\bar{F}^{4},\qquad g^{2}\bar{%
\varepsilon}\bar{D}^{2}\tilde{F}\bar{F}^{3},\qquad g^{2}\bar{\varepsilon}%
\tilde{F}\bar{F}H^{2},\qquad g^{2}\bar{\varepsilon}\bar{D}\tilde{F}\bar{F}H,
\nonumber \\
&&g^{3}\bar{\varepsilon}\tilde{F}\bar{F}^{4},\qquad \frac{g}{\bar{g}}\bar{%
\varepsilon}\hat{D}\bar{D}\bar{F}H.  \label{epsi}
\end{eqnarray}
The first two terms of this list are proportional to each other, which can
be proved using the Bianchi identities. In spite of its appearance, the
first term, which can be written as 
\[
\varepsilon _{\mu \nu \rho \sigma }F_{\mu \nu }\hat{D}^{2}F_{\rho \sigma }, 
\]
does not contain higher time derivatives, since 
\[
\varepsilon _{\bar{\mu}\bar{\nu}\bar{\rho}}\hat{\partial}A_{\bar{\mu}}^{a}%
\hat{\partial}^{2}F_{\bar{\nu}\bar{\rho}}^{a}=g\varepsilon _{\bar{\mu}\bar{%
\nu}\bar{\rho}}f^{abc}\hat{\partial}A_{\bar{\mu}}^{a}\hat{\partial}A_{\bar{%
\nu}}^{b}\hat{\partial}A_{\bar{\rho}}^{c}+\text{ total derivatives.} 
\]
It is also easy to see that no term of the list (\ref{epsi}) affects the
propagators.

The remaining vertices are obtained from the ones listed so far, suppressing
some fields or derivatives, or replacing a fermion with its conjugate (and
then adding the Hermitian conjugate). Finally, we must impose invariance
under $SU(3)\times SU(2)\times U(1)$ (see next section).

The terms $\bar{\varepsilon}\hat{D}^{2}\tilde{F}\bar{F}$, $\bar{\varepsilon}%
\hat{D}\bar{D}\tilde{F}^{2}$ and $g\bar{\varepsilon}\tilde{F}^{3}$ of (\ref
{epsi}) are particular, because they have three $\hat{\partial}$'s, one too
much to fit into the proof given in \cite{LVgauge1suAbar} about the absence
of spurious subdivergences. Fortunately, they can consistently be dropped,
because (\ref{mixec}) ensures that they are not generated back by
renormalization.

Dropping all vertices and quadratic terms that renormalization cannot
generate back we obtain a simplified model. Of course, we cannot drop the
quadratic terms that control the ultraviolet behavior of propagators and the
vertices related to them by covariantization. According to (\ref{mixec})
renormalization potentially generates back only terms of the form 
\[
\bar{g}_{1}^{q+r}\bar{g}_{2}^{2s}\bar{g}_{3}^{t}\hat{D}^{k}\bar{D}^{m}\tilde{%
F}^{q}\bar{F}^{r}\psi ^{2s}H^{t},
\]
which make a much shorter list, precisely the fermionic quadratic terms and
\begin{eqnarray}
&&g^{2}\bar{\varepsilon}\tilde{F}\bar{F},\qquad \bar{g}^{2}H^{2}\bar{\psi}%
\psi ,\qquad (\bar{\psi}\psi )^{2},\qquad \bar{g}\bar{D}H\bar{\psi}\psi
,\qquad g\bar{D}\bar{F}\bar{\psi}\psi ,  \nonumber \\
&&\bar{g}^{2}H\bar{D}^{2}H,\qquad \bar{g}^{4}H^{4},\qquad g^{2}\bar{g}\bar{F}%
^{2}H,\qquad g^{2}\bar{F}\bar{D}^{2}\bar{F},\qquad g^{3}\bar{F}^{3}
\label{shortlist}
\end{eqnarray}
(at $\kappa _{2}=0$). The first term is a total derivative.

There exist other types of Standard Model extensions, actually infinitely
many, since according to (\ref{tutto}) every odd $n$ is allowed (assuming
CPT invariance). However, at present those alternatives appear to be less
interesting than the one singled out here. If we break also CPT\ invariance,
we can have neutrino mass terms like the first of (\ref{cpt}). It is of
course possible to break also rotational invariance, but this does not
change the structure of the theory with respect to the weighted power
counting.

\section{The model}

\setcounter{equation}{0}

In this section we present our model in detail. For practical purposes, we
call it the (Lorentz violating) Standard-Extended Model (SEM).

\paragraph{Gauge-field sector}

The gauge-field lagrangian reads 
\[
\mathcal{L}_{g}=\mathcal{L}_{\text{kin }g}+\mathcal{L}_{\text{int}\ g}+%
\mathcal{L}_{\text{int}\ \varepsilon g}, 
\]
where 
\begin{equation}
\mathcal{L}_{\text{kin }g}=\frac{1}{4}\sum_{G}\left\{ 2F_{\hat{\mu}\bar{\nu}%
}^{G}\eta ^{G}(\bar{\Upsilon})F_{\hat{\mu}\bar{\nu}}^{G}+F_{\bar{\mu}\bar{\nu%
}}^{G}\tau ^{G}(\bar{\Upsilon})F_{\bar{\mu}\bar{\nu}}^{G}+\frac{1}{\Lambda
_{L}^{2}}(D_{\hat{\rho}}F_{\bar{\mu}\bar{\nu}}^{G})\xi ^{G}(\bar{\Upsilon}%
)(D_{\hat{\rho}}F_{\bar{\mu}\bar{\nu}}^{G})\right\} ,  \label{l02}
\end{equation}
and $\sum_{G}$ denotes the sum over the gauge groups $SU(3)$, $SU(2)$ and $%
U(1)$. Moreover, $\bar{\Upsilon}=-\bar{D}^{2}/\Lambda _{L}^{2}$ and 
\[
\eta ^{G}(\bar{\Upsilon})=\sum_{i=0}^{2}\eta _{2-i}^{G}\bar{\Upsilon}%
^{i},\qquad \tau ^{G}(\bar{\Upsilon})=\sum_{i=0}^{4}\tau _{4-i}^{G}\bar{%
\Upsilon}^{i},\qquad \xi ^{G}(\bar{\Upsilon})=\sum_{i=0}^{1}\xi _{1-i}^{G}%
\bar{\Upsilon}^{i}, 
\]
while, symbolically, 
\begin{eqnarray}
\mathcal{L}_{\text{int}\ g} &=&\frac{g\lambda _{3}}{\Lambda _{L}^{2}}\tilde{F%
}^{2}\bar{F}+\frac{g\lambda _{3}^{\prime }}{\Lambda _{L}^{2}}\bar{F}^{3}+g%
\bar{D}^{2}\tilde{F}^{2}\bar{F}+g\hat{D}\bar{D}\tilde{F}\bar{F}^{2}+g\hat{D}%
^{2}\bar{F}^{3}+g^{2}\tilde{F}^{2}\bar{F}^{2}  \nonumber \\
&&+g\bar{D}^{6}\bar{F}^{3}{}+\sum_{r=3}^{4}g^{r-2}\bar{D}^{4}\bar{F}%
^{r}+\sum_{r=3}^{5}g^{r-2}\bar{D}^{2}\bar{F}^{r}{}+\sum_{r=4}^{6}g^{r-2}\bar{%
F}^{r},  \label{l04}
\end{eqnarray}
and 
\begin{eqnarray}
\mathcal{L}_{\text{int}\ \varepsilon g} &=&\bar{\varepsilon}\hat{D}\bar{D}%
^{5}\bar{F}^{2}+\sum_{i=2}^{3}g^{i-1}\bar{\varepsilon}\hat{D}\bar{D}^{3}\bar{%
F}^{i}+\sum_{i=2}^{4}g^{i-2}\bar{\varepsilon}\hat{D}\bar{D}\bar{F}^{i}+\bar{%
\varepsilon}\bar{D}^{6}\tilde{F}\bar{F}  \nonumber \\
&&+\sum_{i=1}^{2}g^{i-1}\bar{\varepsilon}\bar{D}^{4}\tilde{F}\bar{F}%
^{i}+\sum_{i=1}^{3}g^{i-1}\bar{\varepsilon}\bar{D}^{2}\tilde{F}\bar{F}%
^{i}+\sum_{i=1}^{4}g^{i-1}\bar{\varepsilon}\tilde{F}\bar{F}^{i}.
\label{l04eps}
\end{eqnarray}
Here and in the rest of the paper $g$ collectively stands for any of the $%
SU(3)$, $SU(2)$ and $U(1)$ gauge couplings.

\paragraph{Higgs-field sector}

The Higgs-field lagrangian reads 
\[
\mathcal{L}_{H}=\mathcal{L}_{\text{kin}\ H}+\mathcal{L}_{\text{int}\ H}, 
\]
where 
\begin{eqnarray}
\mathcal{L}_{\text{kin}\ H} &=&|\hat{D}_{\hat{\mu}}H|^{2}+\frac{a_{0}}{%
\Lambda _{L}^{4}}|\bar{D}^{2}\bar{D}_{\bar{\mu}}H|^{2}+\frac{a_{1}}{\Lambda
_{L}^{2}}|\bar{D}^{2}H|^{2}+a_{2}|\bar{D}_{\bar{\mu}}H|^{2}+\mu
_{H}^{2}|H|^{2},  \label{h1} \\
\mathcal{L}_{\text{int}\ H} &=&\frac{\lambda _{6}\bar{g}^{4}}{36\Lambda
_{L}^{2}}|H|^{6}+\frac{\lambda _{4}^{(3)}\bar{g}^{2}}{4\Lambda _{L}^{2}}%
|H|^{2}|\bar{D}_{\bar{\mu}}H|^{2}+\frac{\lambda _{4}^{(2)}\bar{g}^{2}}{%
4\Lambda _{L}^{2}}|H^{\dagger }\bar{D}_{\bar{\mu}}H|^{2}  \nonumber \\
&&+\frac{\bar{g}^{2}}{4\Lambda _{L}^{2}}\left[ \lambda _{4}^{(1)}(H^{\dagger
}\bar{D}_{\bar{\mu}}H)^{2}+\text{h.c.}\right] +\frac{\lambda _{4}\bar{g}^{2}%
}{4}|H|^{4},  \label{h2}
\end{eqnarray}
and $a_{i}$, $\lambda _{i}^{(j)}$ and $\mu _{H}$ are constants.

\paragraph{Fermions}

The fermion kinetic terms are 
\begin{equation}
\mathcal{L}_{\text{kin }f}=\sum_{a,b=1}^{3}\sum_{I=1}^{5}\bar{\chi}%
_{I}^{a}\left( \delta ^{ab}\hat{D}\!\!\!\!\slash +\frac{b_{0}^{Iab}}{\Lambda
_{L}^{2}}{\bar{D}\!\!\!\!\slash}\,^{3}+b_{1}^{Iab}\bar{D}\!\!\!\!\slash %
\right) \chi _{I}^{b},  \label{f1}
\end{equation}
where $\chi _{1}^{a}=L^{a}=(\nu _{L}^{a},\ell _{L}^{a})$, $\chi
_{2}^{a}=Q_{L}^{a}=(u_{L}^{a},d_{L}^{a})$, $\chi _{3}^{a}=\ell _{R}^{a}$, $%
\chi _{4}^{a}=u_{R}^{a}$ and $\chi _{5}^{a}=d_{R}^{a}$. Moreover, $\nu
^{a}=(\nu _{e},\nu _{\mu },\nu _{\tau })$, $\ell ^{a}=(e,\mu ,\tau )$, $%
u^{a}=(u,c,t)$ and $d^{a}=(d,s,b)$.

The interactions between fermions and the Higgs field can be divided into
three sets: the usual Yukawa interactions 
\begin{equation}
\mathcal{L}_{\text{Yukawa}}=\bar{g}\sum_{a,b=1}^{3}(Y_{1}^{ab}\bar{L}%
^{ai}\ell _{R}^{b}+Y_{2}^{ab}\bar{u}_{R}^{a}Q_{L}^{bj}\varepsilon
^{ji}+Y_{3}^{ab}\bar{Q}_{L}^{ai}d_{R}^{b})H^{i}+\text{h.c.},  \label{fy}
\end{equation}
where $\bar{g}$ stands for $\bar{g}_{3}$, the vertex (\ref{lh2}), 
\begin{equation}
\mathcal{L}_{LH}=\frac{\bar{g}^{2}}{4\Lambda _{L}}(LH)^{2},  \label{bid}
\end{equation}
and, if $\kappa _{2}=0$, the four fermion interactions, 
\begin{equation}
\mathcal{L}_{4f}\sim \frac{Y_{f}}{\Lambda _{L}^{2}}\bar{\psi}\psi \bar{\psi}%
\psi ,  \label{f4}
\end{equation}
which we do not list here, but can be worked out from ref.s \cite{4f}, and
include the Lorentz violating structures 
\begin{eqnarray}
&&(\bar{\psi}_{1}\psi _{2})(\bar{\psi}_{3}\psi _{4}),\qquad (\bar{\psi}%
_{1}\gamma _{\hat{\mu}}\psi _{2})(\bar{\psi}_{3}\gamma _{\hat{\mu}}\psi
_{4}),\qquad (\bar{\psi}_{1}\gamma _{\bar{\mu}}\psi _{2})(\bar{\psi}%
_{3}\gamma _{\bar{\mu}}\psi _{4}),  \nonumber \\
&&(\bar{\psi}_{1}\sigma _{\hat{\mu}\bar{\nu}}\psi _{2})(\bar{\psi}_{3}\sigma
_{\hat{\mu}\bar{\nu}}\psi _{4}),\qquad (\bar{\psi}_{1}\sigma _{\bar{\mu}\bar{%
\nu}}\psi _{2})(\bar{\psi}_{3}\sigma _{\bar{\mu}\bar{\nu}}\psi _{4}).
\label{ff}
\end{eqnarray}
Here the $\psi _{i}$'s can stand also the conjugate spinors $\psi _{i}^{c}$.
In (\ref{f4}) $Y_{f}$ generically denotes the four fermion couplings.

\paragraph{Matter-gauge-field interactions}

Finally, the lagrangian contains interactions between the Higgs field and
the gauge fields, namely 
\begin{equation}
\mathcal{L}_{\text{int}\ gH}=\left( g\bar{D}^{4}\bar{F}+\sum_{r=1}^{2}g^{r}%
\bar{D}^{2}\bar{F}^{r}+\sum_{r=2}^{3}g^{r}\bar{F}^{r}+g\bar{\varepsilon}\hat{%
D}\bar{D}\bar{F}+g\bar{\varepsilon}\bar{D}^{2}\tilde{F}+g^{2}\bar{\varepsilon%
}\tilde{F}\bar{F}\right) H^{\dagger }H,  \label{gh}
\end{equation}
plus interactions among fermions and gauge fields, 
\begin{equation}
\mathcal{L}_{\text{int}\ gf}=\sum_{I=1}^{5}g\bar{D}\bar{F}\,(\bar{\chi}_{I}%
\bar{\Gamma}\chi _{I}),  \label{fg}
\end{equation}
where $\bar{\Gamma}$ is a matrix $\bar{\gamma}$ or a product of three
matrices $\bar{\gamma}$.

Formulas (\ref{l02}), (\ref{h1}), (\ref{h2}), (\ref{f1}), (\ref{fy}) and (%
\ref{bid}) contain the precise lists of terms, while formulas (\ref{l04}), (%
\ref{l04eps}), (\ref{f4}), (\ref{gh}) and (\ref{fg}) are symbolic, which
means that they contain the basic structures of vertices. Derivatives can
act on the fields and be contracted in all independent ways.

The total lagrangian reads 
\begin{equation}
\mathcal{L}=\mathcal{L}_{g}+\mathcal{L}_{H}+\mathcal{L}_{\text{kin}\ f}+%
\mathcal{L}_{\text{Yukawa}}+\mathcal{L}_{\text{LH}}+\mathcal{L}_{\text{int}\
gH}+\mathcal{L}_{\text{int}\ gf}+\mathcal{L}_{4f}.  \label{totalL}
\end{equation}
The four fermion vertices (\ref{f4}) are the only strictly-renormalizable
interactions, therefore the $Y_{f}$ beta function is proportional to $Y_{f}$%
. Consequently, it is consistent to set $Y_{f}=0$, which gives the solution (%
\ref{sm2}).

\paragraph{Simplified model}

The simplified model is obtained keeping only the terms that are potentially
generated back by renormalization (plus those that we want in any case,
namely (\ref{bid}) and (\ref{f4})). Note that also $\xi $ can be set to
zero, since it is not crucial for the propagator. We have 
\begin{eqnarray}
\mathcal{L}_{\mathrm{simpl}} &=&\frac{1}{4}\sum_{G}\left( 2F_{\hat{\mu}\bar{%
\nu}}^{G}\eta ^{G}(\bar{\Upsilon})F_{\hat{\mu}\bar{\nu}}^{G}+F_{\bar{\mu}%
\bar{\nu}}^{G}\tau ^{G}(\bar{\Upsilon})F_{\bar{\mu}\bar{\nu}}^{G}\right)
+\sum_{a,b=1}^{3}\sum_{I=1}^{5}\bar{\chi}_{I}^{a}\left( \delta ^{ab}\hat{D}%
\!\!\!\!\slash +\frac{b_{0}^{Iab}}{\Lambda _{L}^{2}}{\bar{D}\!\!\!\!\slash}%
\,^{3}+b_{1}^{Iab}\bar{D}\!\!\!\!\slash \right) \chi _{I}^{b}  \nonumber \\
&&+|\hat{D}_{\hat{\mu}}H|^{2}+\frac{a_{0}}{\Lambda _{L}^{4}}|\bar{D}^{2}\bar{%
D}_{\bar{\mu}}H|^{2}+\frac{a_{1}}{\Lambda _{L}^{2}}|\bar{D}^{2}H|^{2}+a_{2}|%
\bar{D}_{\bar{\mu}}H|^{2}+\mu _{H}^{2}|H|^{2}+\frac{\lambda _{4}\bar{g}^{2}}{%
4}|H|^{4}  \nonumber \\
&&+\bar{g}\left( \sum_{a,b=1}^{3}(Y_{1}^{ab}\bar{L}^{ai}\ell
_{R}^{b}+Y_{2}^{ab}\bar{u}_{R}^{a}Q_{L}^{bj}\varepsilon ^{ji}+Y_{3}^{ab}\bar{%
Q}_{L}^{ai}d_{R}^{b})H^{i}+\text{h.c.}\right) +\frac{\bar{g}^{2}}{4\Lambda
_{L}}(LH)^{2}  \nonumber \\
&&+\sum_{I=1}^{5}\frac{1}{\Lambda _{L}^{2}}g\bar{D}\bar{F}\,(\bar{\chi}_{I}%
\bar{\Gamma}\chi _{I})+\frac{Y_{f}}{\Lambda _{L}^{2}}\bar{\psi}\psi \bar{\psi%
}\psi +\frac{g}{\Lambda _{L}^{2}}\bar{F}^{3},  \label{simplL}
\end{eqnarray}
the last three terms being symbolic.

\paragraph{Low-energy Lorentz recovery}

Observe that in this paper $\Lambda _{L}$ denotes the energy scale of
Lorentz violations, while in the literature \cite{strumia} the same symbol
is used, precisely in the same place (\ref{bid}), to denote the energy scale
of lepton number violation. Assuming that $\bar{g}$ and $Y^{ab}$ are of
order 1, we have 
\[
\Lambda _{L}\sim 10^{14}\mathrm{GeV}\text{.} 
\]
The four fermion vertices (\ref{f4}) normally have a different scale, called 
$\Lambda _{B}$ in case they violate baryon number conservation. Present
experimental data give a bound $\Lambda _{B}\gtrsim 10^{15}\mathrm{GeV}$,
not so far from the value of $\Lambda _{L}$.

The low-energy limit of the Standard-Extended Model can be studied taking $%
\Lambda _{L}$ to infinity, which gives the CPT invariant, rotationally
invariant sector of the minimal Standard-Model extension of Colladay and
Kostelecky \cite{kostelecky}, 
\begin{eqnarray}
\mathcal{L}_{\mathrm{lowE}} &=&\sum_{G}\left( \frac{\eta _{2}^{G}}{2}F_{\hat{%
\mu}\bar{\nu}}^{G}F_{\hat{\mu}\bar{\nu}}^{G}+\frac{\tau _{4}^{G}}{4}F_{\bar{%
\mu}\bar{\nu}}^{G}F_{\bar{\mu}\bar{\nu}}^{G}\right)
+\sum_{a,b=1}^{3}\sum_{I=1}^{5}\bar{\chi}_{I}^{a}\left( \delta ^{ab}\hat{D}%
\!\!\!\!\slash +b_{1}^{Iab}\bar{D}\!\!\!\!\slash \right) \chi _{I}^{b} 
\nonumber \\
&&+|\hat{D}_{\hat{\mu}}H|^{2}+a_{2}|\bar{D}_{\bar{\mu}}H|^{2}+\mu
_{H}^{2}|H|^{2}+\frac{\lambda _{4}\bar{g}^{2}}{4}|H|^{4}  \nonumber \\
&&+\bar{g}\left( \sum_{a,b=1}^{3}(Y_{1}^{ab}\bar{L}^{ai}\ell
_{R}^{b}+Y_{2}^{ab}\bar{u}_{R}^{a}Q_{L}^{bj}\varepsilon ^{ji}+Y_{3}^{ab}\bar{%
Q}_{L}^{ai}d_{R}^{b})H^{i}+\text{h.c.}\right) .  \label{lowE}
\end{eqnarray}
Lorentz invariance is recovered in this limit if the couplings $\eta _{2}$, $%
\tau _{4}$, $a_{2}$ and $b_{1}^{I}$ are equal to one at low energies. Four
such conditions can be implemented normalizing the three gauge fields and
suitably rescaling $\bar{x}$ \cite{LVgauge1suA}. Using this freedom, we can
set for example $\eta _{2}=1$ for all gauge fields and, say, $\tau _{4}=1$
for the $U(1)$ gauge field. The Hermitian matrices $b_{1}^{Iab}$ can be
diagonalized by means of unitary transformations $\chi _{I}^{a}\rightarrow
U_{I}^{ab}\chi _{I}^{b}$ (no sum over $I$\ being understood). Let $%
b_{1}^{Ia} $ denote the eigenvalues of $b_{1}^{Iab}$. After such
redefinitions, the Lorentz invariant low-energy surface is parameterized by
the equations 
\begin{equation}
\tau _{4}^{SU(2)}=\tau _{4}^{SU(3)}=a_{2}=b_{1}^{Ia}=1.  \label{ft}
\end{equation}
In general, there is no reason why the low-energy couplings should be
located on this surface. Generically speaking, once Lorentz symmetry is
violated at some energy, renormalization implies that it is violated at all
lower energies. At worst, we have to advocate an appropriate fine tuning.

Call $\delta _{i}$ the differences $\tau _{4}^{SU(2)}-1$, $\tau
_{4}^{SU(3)}-1$, $a_{2}-1$ and $b_{1}^{Ia}-1$. They measure the displacement
from the Lorentz invariant surface in parameter space, normalized so that
the Lorentz surface is $\delta _{i}=0$. Since the surface is RG\ invariant,
the $\delta _{i}$ beta functions are linear combinations of the $\delta $'s.
For $\delta \ll 1$, write $\beta _{\delta }^{i}=c^{ij}\delta ^{j}$, $c^{ij}$
being a matrix depending on the other couplings. Solving the RG equations we
find $\delta ^{i}(\mu )=(\mu /\mu ^{\prime })^{c^{ij}}\delta ^{j}(\mu
^{\prime })$. Therefore, the Lorentz surface is RG stable if the matrix $%
c^{ij}$ is positive definite. Evidence that it is so was given in ref.s \cite
{nielsen}, for CPT\ invariant Lorentz violations. There is also evidence
that CPT\ violating terms exhibit the opposite behavior \cite{colladay}.

Observe that (\ref{lh2}) is the unique lagrangian term of dimension five.
Kinetic terms of dimension five are not allowed, because the unique
candidates, $\bar{\psi}_{L}\bar{D}^{2}\psi _{R}$, $\bar{\psi}_{L}^{c}\bar{D}%
^{2}\psi _{L}$, and so on, are excluded by $SU(2)\times U(1)$ invariance.
Candidate dimension-five vertices, such as $\bar{F}_{\mu \nu }\bar{\psi}%
_{L}\sigma ^{\bar{\mu}\bar{\nu}}\psi _{R}$, $\bar{D}\bar{\psi}_{L}\bar{\gamma%
}\psi _{L}H$, $\bar{F}_{\mu \nu }\bar{\psi}_{L}^{c}\sigma ^{\bar{\mu}\bar{\nu%
}}\psi _{L}$, $\bar{D}\bar{\psi}_{R}^{c}\bar{\gamma}\psi _{L}H$, are again
forbidden by $SU(2)\times U(1)$ invariance. Therefore, if we assume the
low-energy fine-tuning (\ref{ft}), then not only the power-counting
renormalizable subsector, but also the dimension-five subsector are Lorentz
invariant. The Lorentz violations predicted by the SEM can be tested in
experiments that are sensitive to the effects of the dimension-six subsector
(or the $\delta ^{i}$-running).

Energies of the order of $10^{14}\mathrm{GeV}$ are out of reach in present
high-energy experiments, so, if we want to distinguish the SEM from a
see-saw mechanism, it is necessary to develop methods to amplify small
effects. One way is to consider situations where it is possible to observe a
huge number of copies of a system at the same time. Examples are the
searches for proton decay and double beta decay. The double beta decay
admits a neutrinoless version, which, if observed, can prove the existence
of Majorana masses. The SEM does not predict large modifications to such
phenomena. The four fermion vertices involved in proton decay are in
principle sensitive to Lorentz violations (see formula (\ref{ff})), but
measuring one quantity alone (the proton lifetime) will not be enough to
discriminate the SEM from other proposals. Nevertheless, the advantage of
having a comprehensive model is that it might have implications that are not
apparent at first sight. Further work is required to fit a feasible
experimental setting to the predictions of the SEM.

Finally, it is worth emphasizing that although other Lorentz violating
extensions of the Standard Model can be constructed with $n=5$, $7$, etc.,
the SEM has the right weighted dimension (\dj $=2$) to accommodate a Lorentz
violating version of quantum gravity. Indeed, in the framework of ordinary
power counting gravity is renormalizable in $d=2$ (where however it is
trivial), so it is reasonable to expect that weighted power counting can
renormalize its Lorentz violating version in \dj $=2$.

\section{Anomalies and anomaly cancellation}

\setcounter{equation}{0}

In this section we prove that the gauge anomalies of the SEM coincide with
those of the Standard Model, therefore they cancel out to all orders. We use
the Batalin-Vilkovisky formalism \cite{batalin}. The fields are collectively
denoted by $\Phi ^{i}=(\hat{A}_{\hat{\mu}}^{a},\bar{A}_{\bar{\mu}}^{a},%
\overline{C}^{a},C^{a},B^{a},\psi ,\bar{\psi},\varphi )$. Add BRST sources $%
K_{i}=(\hat{K}_{a}^{\hat{\mu}},\bar{K}_{a}^{\bar{\mu}},K_{\overline{C}%
}^{a},K_{C}^{a},K_{B}^{a},K_{\psi },K_{\bar{\psi}},K_{\varphi })$ for every
field $\Phi ^{i}$ and extend the action $\mathcal{S}(\Phi )$ (the integral
of (\ref{totalL}) or (\ref{simplL})) to 
\begin{equation}
\Sigma (\Phi ,K)=\mathcal{S}(\Phi )-\int \mathrm{d}^{d}x\sum_{i}\left( s\Phi
^{i}\right) K_{i},  \label{acca}
\end{equation}
Define the antiparenthesis 
\begin{equation}
(X,Y)=\int \mathrm{d}^{d}x\left\{ \frac{\delta _{r}X}{\delta \Phi ^{i}(x)}%
\frac{\delta _{l}Y}{\delta K_{i}(x)}-\frac{\delta _{r}X}{\delta K_{i}(x)}%
\frac{\delta _{l}Y}{\delta \Phi ^{i}(x)}\right\} .  \label{antipar}
\end{equation}
BRST\ invariance is generalized to the identity 
\begin{equation}
(\Sigma ,\Sigma )=0,  \label{nil}
\end{equation}
which is a straightforward consequence of (\ref{acca}), the BRST-invariance
of $\mathcal{S}$ and the nilpotency of $s$. If the regularization preserves
the BRST invariance of the functional integration measure, which we assume
here (see below for details), and equation (\ref{nil}) holds at the
regularized level, then we have also 
\begin{equation}
(\Gamma ,\Gamma )=0,  \label{milpo}
\end{equation}
where $\Gamma $ is the generating functional of one-particle irreducible
Green functions. If (\ref{nil}) does not hold at the regularized level, (\ref
{milpo}) can be violated. Gauge anomalies are collected in the functional 
\begin{equation}
\mathcal{A}(\Phi ,K)\equiv (\Gamma ,\Gamma )=\langle (\Sigma ,\Sigma
)\rangle .  \label{anom}
\end{equation}
Since $(X,(X,X))\equiv 0$ for every functional $X$, we have 
\begin{equation}
(\Gamma ,\mathcal{A})=0,  \label{coho}
\end{equation}
which is the Wess-Zumino consistency condition \cite{wz}, written in BRST
language. We also know that the one-loop contribution $\mathcal{A}_{\text{%
1-loop}}(\Phi ,K)$ to the anomaly is a local functional, precisely the
integral of a local function of dimension five, weight 3 and ghost number 1.

\paragraph{Regularization}

The most convenient regularization framework to study anomalies is \cite
{renoanom} a combination of a (Lorentz violating) higher-derivative
regularization \textsl{\`{a} la} Slavnov \cite{slavnov} with the
dimensional-regularization technique. We recall that $\hat{d}$ and $\bar{d}$
have to be analytically continued independently, say to $\hat{d}-\varepsilon
_{1}$ and $\bar{d}-\varepsilon _{2}$, respectively \cite{renolor,confnolor}.
The tensor $\varepsilon _{\bar{\mu}\bar{\nu}\bar{\rho}}$ and the matrix $%
\gamma _{5}$ are defined according to the 't Hooft-Veltman prescription \cite
{collins2}.

The (partial) higher-derivative regularization is defined as follows. The
pure-gauge sector is extended replacing $\mathcal{L}_{Q}$ with 
\begin{equation}
\mathcal{L}_{Q\text{-HD}}=\frac{1}{4}\left\{ 2F_{\hat{\mu}\bar{\nu}}\tilde{Q}%
_{M}(\hat{D}^{2},\bar{D}^{2})F_{\hat{\mu}\bar{\nu}}+F_{\bar{\mu}\bar{\nu}}%
\bar{Q}_{M}(\hat{D}^{2},\bar{D}^{2})F_{\bar{\mu}\bar{\nu}}\right\} =\mathcal{%
L}_{Q}+\mathcal{O}(1/\Lambda ^{2}),  \label{spure}
\end{equation}
where the $Q_{M}$'s are polynomials of degree $M>2(n-1)=4$ and $\Lambda $
denotes the cutoff. The gauge-fixing $\mathcal{G}^{a}$ is modified as 
\[
\mathcal{G}_{\text{HD}}^{a}=\hat{G}_{M}(\hat{\partial}^{2},\bar{\partial}%
^{2})\hat{\partial}\cdot \hat{A}^{a}+\bar{G}_{M}(\hat{\partial}^{2},\bar{%
\partial}^{2})\bar{\partial}\cdot \bar{A}^{a}=\mathcal{G}^{a}+\mathcal{O}%
(1/\Lambda ^{2}), 
\]
where the $G_{M}$'s are polynomials of degree $M$. The gauge-field and ghost
propagators fall off as $1/k^{2M+2}$ for large $k$. With suitable positivity
conditions on the coefficients in $Q_{M}$ and $G_{M}$ the propagators are
regular (but violate unitarity). The Higgs field is regularized replacing $%
\mathcal{L}_{\text{kin}\ H}$ with 
\begin{equation}
\mathcal{L}_{\text{kin}\ H\text{-HD}}=H^{\dagger }Q_{M,H}(\hat{D}^{2},\bar{D}%
^{2})H=\mathcal{L}_{\text{kin}\ H}+\mathcal{O}(1/\Lambda ^{2}),  \label{hr}
\end{equation}
where $Q_{M,H}$ is a polynomial of degree $M+1$. Finally, the fermions are
regularized as 
\begin{equation}
\mathcal{L}_{\text{kin }f\text{-HD}}=\sum_{a,b=1}^{3}\sum_{I=1}^{5}\bar{\chi}%
_{I}^{a}\left[ \hat{Q}_{M,f}^{ab}(\hat{D}^{2},\bar{D}^{2})\hat{D}\!\!\!\!%
\slash +\bar{Q}_{M,f}^{ab}(\hat{D}^{2},\bar{D}^{2}){\bar{D}\!\!\!\!\slash}%
)\right] \chi _{I}^{b}=\mathcal{L}_{\text{kin }f}+\mathcal{O}(1/\Lambda ),
\label{f1r}
\end{equation}
where the $Q_{M,f}^{ab}$'s are polynomials of degree $M$.

So far, BRST invariance is manifestly preserved (also at $\varepsilon
_{1,2}\neq 0$). However, since fermions have definite chiralities, the
quadratic terms of (\ref{f1}) or (\ref{f1r}) dot not give good propagators
at $\varepsilon _{1,2}\neq 0$. We introduce extra fermions $\tilde{\chi}%
_{I}^{a}$ with chiralities opposite to those of the $\chi _{I}^{a}$'s and
the same $SU(3)\times SU(2)\times U(1)$-representations. Then, we collect $%
\chi _{I}^{a}$ and $\tilde{\chi}_{I}^{a}$ into Dirac fermions $\psi _{I}^{a}$%
, and replace the \textit{free} fermionic lagrangian with 
\begin{equation}
\sum_{a,b=1}^{3}\sum_{I=1}^{5}\bar{\psi}_{I}^{a}\left[ \hat{Q}_{M,f}^{ab}(%
\hat{\partial}^{2},\bar{\partial}^{2})\hat{\partial}\!\!\!\slash +\bar{Q}%
_{M,f}^{ab}(\hat{\partial}^{2},\bar{\partial}^{2}){\bar{\partial}\!\!\!\slash%
})\right] \psi _{I}^{b}.  \label{chira}
\end{equation}
When $\varepsilon _{1}=\varepsilon _{2}=0$ the fields $\tilde{\chi}_{I}^{a}$
are free and decouple, but when $\varepsilon _{1}$ and $\varepsilon _{2}$
are nonzero they couple with the $\chi _{I}^{a}$'s in ``evanescent'' ways 
\cite{collins}, namely by terms that formally disappear in the physical
space and time dimensions. We know that there exists a subtraction scheme
where the $\tilde{\chi}_{I}^{a}$'s stay decoupled also in $\Gamma $, because
evanescent operators do not mix into the physical ones \cite{evane}. Now,
the modification (\ref{chira}) preserves the global symmetries, but not the
local ones, which can be anomalous. In particular, the violation of $(\Sigma
,\Sigma )=0$ comes only from the new terms contained in (\ref{chira}) and it
is proportional to $\partial C$.

Finally, the physical SEM is defined first renormalizing the
higher-derivative theory (which means that $\Lambda $ is treated as an
ordinary, finite parameter), then taking the limits $\varepsilon
_{1,2}\rightarrow 0$, then renormalizing the $\Lambda $-divergences, and
finally taking the limit $\Lambda \rightarrow \infty $.

It is convenient to introduce tilded quantities as follows: 
\[
\tilde{\Phi}^{i}=\frac{\Phi ^{i}}{\Lambda ^{M}},\qquad \tilde{K}_{i}=\Lambda
^{M}K_{i},\qquad \tilde{g}=g\Lambda ^{M},\qquad \widetilde{\bar{g}}_{2}=\bar{%
g}_{2}\Lambda ^{M},\qquad \widetilde{\bar{g}}_{3}=\bar{g}_{3}\Lambda ^{M}. 
\]
Note that this map is a canonical transformation combined with a coupling
redefinition. At finite $\Lambda $ the ultraviolet behavior of the
higher-derivative theory is governed by ordinary power counting. In the
tilded parametrization the theory has the form (\ref{mixed}), is polynomial
in $\Lambda $ and super-renormalizable. Every parameter not shown explicitly
in (\ref{mixed}) is either of non-negative dimension and $\Lambda $-$\Lambda
_{L}$-independent, or has the form $\lambda \Lambda ^{m}/\Lambda
_{L}^{m^{\prime }}$, where $\lambda $ has a non-negative dimension and $%
m>m^{\prime }$. Differentiating propagators and Feynman diagrams with
respect to $\Lambda $ improves their ultraviolet behaviors.

\paragraph{Anomaly cancellation}

In Feynman diagrams every external $\tilde{K}$-leg is multiplied by a factor 
$\tilde{g}$. This follows from the structure of the vertices contained in $%
\Sigma $. Then, because of (\ref{mixec}) the one-loop counterterms have the
weight structure 
\begin{equation}
\Delta _{1}\mathcal{L}(\tilde{g}\tilde{A},\tilde{g}\widetilde{\bar{C}},%
\tilde{g}\tilde{C},\widetilde{\bar{g}}_{2}\tilde{\psi},\widetilde{\bar{g}}%
_{3}\tilde{\varphi},\tilde{g}\tilde{K}_{i}),  \label{mixec2}
\end{equation}
while at $L$-loops there are $2L-2$ additional factors of $\widetilde{g}$-$%
\widetilde{\bar{g}}_{2}$-$\widetilde{\bar{g}}_{3}$. Observe that the
dimensions of $\tilde{g}\tilde{A}$, $\tilde{g}\widetilde{\bar{C}}$, $\tilde{g%
}\tilde{C}$, $\widetilde{\bar{g}}_{2}\tilde{\psi}$ and $\widetilde{\bar{g}}%
_{3}\tilde{\varphi}$ are $M$-independent, while those of $\tilde{g}\tilde{K}%
_{i}$, as well as those of $\tilde{g}$, $\tilde{g}_{2}$, $\tilde{g}_{3}$ can
be made arbitrarily large choosing $M$ appropriately. Consequently, for
sufficiently large $M$ there exist no divergent diagram beyond one loop (at
finite $\Lambda $) and no divergent diagram with $K$ external legs. The
dimensional technique regularizes the few divergent diagrams of the
higher-derivative theory.

The one-loop anomaly $\mathcal{A}_{\text{1-loop}}$ is the integral of a
local function of dimension five. If $M$ is large enough, $\mathcal{A}_{%
\text{1-loop}}$ must be $K$-independent. Moreover, $\Sigma $ depends on the
antighosts $\overline{C}$ only via the combinations $K_{a}^{\hat{\mu}}+\hat{G%
}_{M}\partial ^{\hat{\mu}}\overline{C}^{a}$ and $K_{a}^{\bar{\mu}}+\bar{G}%
_{M}\partial ^{\bar{\mu}}\overline{C}^{a}$, so the same must be true of $%
\mathcal{A}$. We conclude that $\mathcal{A}_{\text{1-loop}}$ is also $%
\overline{C}$-independent. By ghost number conservation, it must be linear
in $C$. We can write 
\begin{equation}
\mathcal{A}_{\text{1-loop}}=\int \mathrm{d}^{4}x\ \tilde{g}\tilde{C}^{a}%
\mathcal{A}_{a}(\tilde{g}\tilde{A},\widetilde{\bar{g}}_{2}\tilde{\psi},%
\widetilde{\bar{g}}_{3}\tilde{\varphi}),  \label{line}
\end{equation}
where $\mathcal{A}_{a}$ is a local function of dimension four.

We can do even better. Replace the polynomials $Q_{M,H}$ and $Q_{M,f}^{ab}$
in (\ref{hr}), (\ref{f1r}) and (\ref{chira}) with polynomials $Q_{M^{\prime
},H}$ and $Q_{M^{\prime },f}^{ab}$ of degrees $M^{\prime }+1$ and $M^{\prime
}$, respectively, where $M^{\prime }$ is such that 
\begin{equation}
\frac{2M+1}{3}<M^{\prime }<M.  \label{left}
\end{equation}
Since $M>4$, this condition admits solutions. Then, define 
\[
\tilde{\psi}^{\prime }=\frac{\psi }{\Lambda ^{M^{\prime }}},\qquad \tilde{%
\varphi}^{\prime }=\frac{\varphi }{\Lambda ^{M^{\prime }}},\qquad \tilde{K}%
_{\psi }^{\prime }=\Lambda ^{M^{\prime }}K_{\psi },\qquad \tilde{K}_{\varphi
}^{\prime }=\Lambda ^{M^{\prime }}K_{\varphi }, 
\]
and leave all other tilded definitions unmodified. It is easy to check that
the action is still super-renormalizable, of the form (\ref{mixed}) and
polynomial in $\Lambda $. In particular, the left inequality of (\ref{left})
still ensures that in the tilded (\ref{mixed}) parametrization every
parameter not shown exlicitly in (\ref{mixed}) is either of non-negative
dimension and $\Lambda $-$\Lambda _{L}$-independent, or has the form $%
\lambda \Lambda ^{m}/\Lambda _{L}^{m^{\prime }}$, where $\lambda $ has a
non-negative dimension and $m>m^{\prime }$. Again, differentiating
propagators and Feynman diagrams with respect to $\Lambda $ improves their
ultraviolet behaviors.

What we have gained is that now also the dimensions of $\widetilde{\bar{g}}%
_{2}\tilde{\psi}^{\prime }$ and $\widetilde{\bar{g}}_{3}\tilde{\varphi}%
^{\prime }$ can be arbitrarily large, if $M$ is large enough and $M^{\prime
} $ is chosen appropriately. Then counterterms and the one-loop anomaly do
not depend on matter fields. We can write 
\begin{equation}
\mathcal{A}_{\text{1-loop}}=\int \mathrm{d}^{4}x\ \tilde{g}\tilde{C}^{a}%
\mathcal{A}_{a}^{\prime }(\tilde{g}\tilde{A})=\int \mathrm{d}^{4}x\ gC^{a}%
\mathcal{A}_{a}^{\prime }(gA).  \label{line2}
\end{equation}

Now, assume that the SEM\ one-loop anomaly vanishes at finite $\Lambda $.
This result is proved below. Then locality and the structure (\ref{line2})
are inherited by the two-loop anomaly $\mathcal{A}_{\text{2-loop}}$, with $2$
additional factors of $\widetilde{g}$-$\widetilde{\bar{g}}_{2}$-$\widetilde{%
\bar{g}}_{3}$ in front. If $M$ is sufficiently large power counting implies $%
\mathcal{A}_{\text{2-loop}}=0$. Iterating the argument we find that the
anomaly cancels identically at finite $\Lambda $ (and $\varepsilon _{1,2}=0$%
), namely $(\Gamma ,\Gamma )\equiv 0$. Therefore the higher-derivative
theory has no gauge anomaly.

At this point we can take $\Lambda $ large, subtract the $\Lambda $%
-divergences and finally send $\Lambda $ to infinity, which gives the SEM.
Since $(\Gamma ,\Gamma )$ vanishes for arbitrary finite $\Lambda $, the $%
\Lambda $-divergences are BRST\ invariant and can be subtracted in a BRST\
invariant way, therefore preserving the identity $(\Gamma ,\Gamma )=0$ and
the anomaly cancellation. In practice, the dimensional/higher-derivative
framework is a manifestly BRST\ invariant regularization of the SEM. We
conclude that the Standard-Extended Model has no gauge anomalies.

After sending $\Lambda $ to infinity we can also take the limit $\Lambda
_{L}\rightarrow \infty $, which gives the low-energy model (\ref{lowE}).
Again, the anomaly cancellation survives the limit, because it holds at
arbitrary finite $\Lambda _{L}$. Therefore, the model (\ref{lowE}) has no
gauge anomalies. Since the Standard Model is just the model (\ref{lowE})
with the relations (\ref{ft}), the same argument provides also an
alternative proof of the cancellation of gauge anomalies in the Standard
Model to all orders.

\paragraph{One-loop cancellation}

It remains to prove that the one-loop anomalies cancel at finite $\Lambda $.
Using (\ref{line2}) and taking the first order of equation (\ref{coho}) we
get 
\begin{equation}
s\mathcal{A}_{\text{1-loop}}=0.  \label{coho3}
\end{equation}
This problem can be solved in a cohomological sense, namely up to
functionals that can be written as BRST\ variations of other local
functionals. Indeed, those types of contributions to $\mathcal{A}_{\text{%
1-loop}}$ can be subtracted away with a simple $\Sigma $-redefinition. The
function $\mathcal{A}_{a}^{\prime }(gA)$ is a sum of terms of dimensions $%
\leq 4$. The non-trivial structures of $\mathcal{A}_{\text{1-loop}}$ can be
studied directly. It is easy to show that there exist no non-trivial
structures of dimension $<4$, so $\mathcal{A}_{a}$ is a linear combination
of terms with dimension exactly equal to 4. The coefficients in front of
such terms are dimensionless, therefore they must be $\Lambda $- and $%
\Lambda _{L}$-independent. Indeed, we know that the higher-derivative
theory, in the tilded (\ref{mixed}) parametrization, has only parameters
that are of non-negative dimension and $\Lambda $-$\Lambda _{L}$%
-independent, or have the form $\lambda \Lambda ^{m}/\Lambda _{L}^{m^{\prime
}}$, where $\lambda $ has a non-negative dimension and $m>m^{\prime }$. So,
we are free to compute the one-loop anomaly in the limit $\Lambda
\rightarrow \infty $, which gives the SEM, but also in the limit $\Lambda
\rightarrow \infty $ followed by the limit $\Lambda _{L}\rightarrow \infty $%
, which gives (\ref{lowE}). We conclude that the one-loop anomalies of the
higher-derivative theory coincide with those of the SEM and those of (\ref
{lowE}).

Now we prove that the anomalies of (\ref{lowE}) coincide with those of the
Standard Model, therefore they vanish. The one-loop anomaly of (\ref{lowE})
is the sum of separate contributions due to fermions with actions 
\[
\int \mathrm{d}^{4}x\sum_{a,b=1}^{3}\bar{\chi}_{I}^{a}\left( \delta ^{ab}%
\hat{D}\!\!\!\!\slash +b_{1}^{Iab}\bar{D}\!\!\!\!\slash \right) \chi
_{I}^{b}, 
\]
where no sum over $I$ is understood. The matrices $b_{1}^{Iab}$ are
Hermitian and can be diagonalized with unitary transformations $\chi
_{I}^{a}\rightarrow U_{I}^{ab}\chi _{I}^{b}$, which further reduces $%
\mathcal{A}_{\text{1-loop}}^{Ia}$ to a sum of contributions $\mathcal{A}_{%
\text{1-loop}}^{Ia}$ due to 
\begin{equation}
\int \mathrm{d}^{4}x\ \bar{\chi}_{I}^{a}\left( \hat{D}\!\!\!\!\slash %
+b_{1}^{Ia}\bar{D}\!\!\!\!\slash \right) \chi _{I}^{a},  \label{anoo}
\end{equation}
where no sums over $I$ and $a$ are understood. Finally, the anomaly of (\ref
{anoo}) has the usual value, because it does not depend on $b_{1}^{Ia}$. A
quick way to prove this is to note that $b_{1}^{Ia}$ can be reabsorbed into
the redefinitions 
\begin{equation}
\bar{x}\rightarrow \bar{x}b_{1}^{Ia},\qquad \bar{A}\rightarrow \frac{\bar{A}%
}{b_{1}^{Ia}},\qquad \chi _{I}^{a}\rightarrow \chi
_{I}^{a}(b_{1}^{Ia})^{-3/2},  \label{per}
\end{equation}
after which (\ref{anoo}) becomes fully Lorentz invariant. After this
redefinition $\mathcal{A}_{\text{1-loop}}^{Ia}$ has the usual Bardeen
structure 
\begin{equation}
c^{Ia}g^{3}\varepsilon _{\mu \nu \rho \sigma }\int \mathrm{d}^{4}x\ \mathrm{%
Tr}\left[ \partial _{\mu }C\left( A_{\nu }\partial _{\rho }A_{\sigma }-\frac{%
g}{2}A_{\nu }A_{\rho }A_{\sigma }\right) \right] .  \label{a1loop}
\end{equation}
Now, since (\ref{a1loop}) is invariant under (\ref{per}), it is safe to undo
the redefinition (\ref{per}). Thus, the anomaly $\mathcal{A}_{\text{1-loop}%
}^{Ia}$ of (\ref{anoo}) is (\ref{a1loop}), with $b_{1}^{Ia}$-independent
coefficients $c^{Ia}$. A different analysis of one-loop anomalies in
theories (\ref{anoo}), leading to the same result, was performed in ref. 
\cite{arias}. We conclude that the one-loop anomalies of (\ref{lowE})
coincide with those of the Standard Model, therefore they vanish.

\paragraph{Summary}

Summarizing, the one-loop anomalies of the SEM coincide with those of the
Standard Model, so they cancel. Since they cancel at one loop, there exists
a subtraction scheme where they cancel to all orders. The
dimensional/higher-derivative regularization framework described above
selects the right scheme automatically. Identical arguments and conclusions
apply to the Standard Model and the Lorentz violating Standard Model
extension (\ref{lowE}). Observe that we have not used the assumption (\ref
{ft}) that Lorentz invariance is recovered at low energies. Our argument,
which makes no use of complicated cohomological theorems, provides also a
general and economic proof of the Adler-Bardeen theorem \cite{adler2}. Gauge
anomalies and their cancellation are in some sense universal properties,
since they are not affected by Lorentz violations and radiative corrections.

\section{Conclusions}

\setcounter{equation}{0}

In this paper we have studied the Lorentz violating extensions of the
Standard Model that are renormalizable by weighted power counting. The
theories contain higher space derivatives, but are arranged so that no
counterterms with higher time derivatives are generated, which ensures
perturbative unitarity. Spacetime is split into time and space.

We have searched for ``interesting'' extensions of the Standard Model,
namely models that can, at least, renormalize two scalar-two fermion
vertices, and therefore give masses to the (left-handed) neutrinos without
the need to introduce right-handed neutrinos, nor other extra fields, and
without violating CPT. We have found that the simplest model with such
properties can contain also four fermion interactions, and therefore
describe proton decay. Finally, the cancellation of anomalies is inherited
from the one of the Standard Model. Our model is predictive and offers a new
scenario for the physics beyond the Standard Model.

If we accept that Lorentz invariance is violated at high energies there
remains to explain why it should be recovered at low energies, since
generically renormalization make the couplings run independently. It is of
course possible to restore Lorentz invariance at low energies by means of a
fine tuning, which would be easier to justify if the Lorentz invariant
surface were infrared stable.

Why should we believe that Lorentz symmetry might not be exact at very high
energies? One reason is that the set of renormalizable theories is
considerably larger once the assumption of Lorentz invariance is relaxed.
Moreover, if CPT\ is a symmetry of Nature, then the Standard Model violates
both parity and time reversal. The violations of P and T \textit{are}
Lorentz violations, because they break the large Lorentz group into the
restricted Lorentz group. Maybe they are indications that at higher energies
also the restricted Lorentz group is broken. The smallness of the T\
violation with respect to the P violation could be a sign of ``hierarchy''
among the various types of Lorentz violations. Neutrino masses could be a
further sign of Lorentz violation. Then the scale of Lorentz violation would
be $\Lambda _{L}\sim 10^{14}\mathrm{GeV}$.

\vskip 20truept \noindent {\Large \textbf{Acknowledgments}}

\vskip 10truept

I am grateful to V. Cavasinni, D. Comelli, E. Guadagnini, F. Nesti, G.
Paffuti, L. Pilo and A. Strumia for useful discussions. I thank the referee
for pointing out relevant references.

\vskip 20truept \noindent {\Large \textbf{Appendix: Gauge-field propagator}}

\vskip 10truept

\renewcommand{\theequation}{A.\arabic{equation}} \setcounter{equation}{0}

The (Euclidean) propagator at $\hat{d}=1$ in the ``Feynman gauge'' $\lambda
=1$, $\zeta =\eta $, reads \cite{LVgauge1suA} 
\begin{equation}
\langle A(k)\ A(-k)\rangle =\left( 
\begin{array}{cc}
\langle \hat{A}\hat{A}\rangle & \langle \hat{A}\bar{A}\rangle \\ 
\langle \bar{A}\hat{A}\rangle & \langle \bar{A}\bar{A}\rangle
\end{array}
\right) =\left( 
\begin{array}{cc}
u\hat{\delta} & 0 \\ 
0 & v\bar{\delta}+t\bar{k}\bar{k}
\end{array}
\right) ,  \label{pros}
\end{equation}
where 
\[
u=\frac{1}{D(1,\eta )},\qquad v=\frac{1}{D(\tilde{\eta},\tau )},\qquad t=%
\frac{\tilde{\tau}-\eta ^{2}}{\eta D(\tilde{\eta},\tau )D(1,\eta )}. 
\]
Here 
\[
D(x,y)\equiv x\hat{k}^{2}+y\bar{k}^{2},\qquad \tilde{\eta}=\eta +\frac{\bar{k%
}^{2}}{\Lambda _{L}^{2}}\xi ,\qquad \tilde{\tau}=\tau +\frac{\hat{k}^{2}}{%
\Lambda _{L}^{2}}\xi , 
\]
and now $\eta $, $\tau $ and $\xi $, as well as $x$ and $y$, are functions
of $\bar{k}^{2}/\Lambda _{L}^{2}$. The ghost propagator is $1/D(1,\eta )$.

The physical degrees of freedom can be read in the ``Coulomb'' gauge $\bar{%
\partial}\cdot \bar{A}^{a}=0$, where 
\[
\langle \hat{A}\hat{A}\rangle =\frac{1}{\eta \bar{k}^{2}},\qquad \langle 
\hat{A}\bar{A}\rangle =0,\qquad \langle \bar{A}\bar{A}\rangle =\frac{1}{D(%
\tilde{\eta},\tau )}\left( \bar{\delta}-\frac{\bar{k}\bar{k}}{\bar{k}^{2}}%
\right) , 
\]
so the dispersion relation is (\ref{disp}). The ghosts are non-propagating
in this gauge.

The ``spurious subdivergences'' are the UV divergences of the subintegrals
over $\hat{k}$ or $\bar{k}$. To ensure that those are automatically
subtracted, the propagators must behave correctly not only in the compelte $%
\hat{k}$-$\bar{k}$ integrals, but also in the subintegrals where some $\hat{k%
}$ and/or $\bar{k}$ integrations are missing. The propagators (\ref{pros})
behave correctly for $\bar{k}\rightarrow \infty $, and all of them but $%
\langle \bar{A}\bar{A}\rangle $ behave correctly also for $\hat{k}%
\rightarrow \infty $. Instead, the propagator $\langle \bar{A}\bar{A}\rangle 
$ behaves like $1/\hat{k}^{2}$ for $\hat{k}\rightarrow \infty $, which is
not enough. Therefore, the subintegrals must be studied more closely. When $%
\hat{d}=1$, $d=$even, $n=$odd and other restrictions are fulfilled they can
be proved to be convergent \cite{LVgauge1suA,LVgauge1suAbar}. When those
conditions are not fulfilled, or when the spacetime manifold $M$ is split
into the product of more than two subfactors, subdivergences are present, in
general, and it is not known how to treat them.

\end{document}